# Contribution to the modeling of solar spicules

E. Tavabi[a, b, c, 1], S. Koutchmy[c] and A. Ajabshirizadeh[b, d]

[a] Payame Noor University (PNU), Zanjan, Iran

[b] Research Institute for Astronomy & Astrophysics of Maragha (RIAAM), 55134-441 Maragha, Iran

[c] Institut d'Astrophysique de Paris, UMR 7095, CNRS & UPMC, 98 Bis Bd Arago, F-75014 Paris, France

[d] Department of Theoretical Physics and Astrophysics, Tabriz University, 51664 Tabriz, Iran

E-mail addresses: tavabi@iap.fr (E. Tavabi), koutchmy@iap.fr (S. Koutchmy), a-adjab@tabrizu.ac.ir (A. Ajabshirizadeh).

**Abstract**

Solar limb and disc spicule quasi- periodic motions have been reported for a long time, strongly suggesting that they are oscillating. In order to clear up the origin and possibly explain some solar limb and disc spicule quasi-periodic recurrences produced by overlapping effects, we present a simulation model assuming quasi- random positions of spicules. We also allow a set number of spicules with different physical properties (such as: height, lifetime and tilt angle as shown by an individual spicule) occurring randomly.

Results of simulations made with three different spatial resolutions of the corresponding frames and also for different number density of spicules, are analyzed. The wavelet time/frequency method is used to obtain the exact period of spicule visibility. Results are compared with observations of the chromosphere from i/ the Transition Region and Coronal Explorer (TRACE) filtergrams taken at 1600 angstrom, ii/ the Solar Optical Telescope (SOT) of Hinode taken in the Ca II H-line and iii/ the Sac-Peak Dunn's VTT taken in Hα line. Our results suggest the need to be cautious when interpreting apparent oscillations seen in spicule image sequences when overlapping is present, i.e.; when the spatial resolution is not enough to resolve individual components of spicules.

---

[1] Corresponding author: E. Tavabi is with Payame Noor University (PNU), 45138-69987 Zanjan, Iran,
Fax: +9824142487120
Telephone: +982414224062

### 1- Introduction

Spicules are an important extended rather cool structure between the solar surface and the corona, partly filling the space occupied by the chromosphere and surrounded by a thin transition region. Their origin, formation and dynamical properties are still mysterious.

They look like jet-like structures ejected from a region (sometimes called a rosette) situated near the supergranulation boundaries and they are one of the most important and dramatic dynamic phenomena in the Transition Region (TR) and the chromosphere. The mechanism of spicule formation and evolution is not well understood (for the mechanism of formation, see the review of Sterling, 2000 and also Lorrain and Koutchmy 1996; Filippov et al. 2007).

The investigation of solar spicules is necessary to understand the TR and the coronal heating (Kudoh and Shibata, 1999). The typical lifetime of spicules is 5-15 min. and the erupted speed of spicules from the low chromosphere layers is about 25-50 $km/s$. They usually reach heights of 5000-10,000km before fading out of view or fall back towards the solar surface. Their smallest widths are only 100- 200 $km$ (Tavabi et al. 2011), Tavabi et al. found that indeed spicules show a whole range of diameters, including unresolved "interacting spicules" (I-S), depending of the definition chosen to characterize this ubiquitous dynamical phenomenon occurring into a low coronal surrounding. A definite signature in the 0.18 to 0.25 Mm range exists, corresponding to the occurrence of the newly discovered type II spicules and, even more impressively, large Fourier amplitudes are observed in the 0.3 to the 1.2 Mm range of diameters and spacing. They are observed in many optical lines, such as *Hα, Hβ, H & K of Ca II* and EUV. We want to point out that a series of papers was devoted to *Hα* and to *He*+ spicules by Georgakilas et al. (1999) and Christopoulou et al. (2001) and an exhaustive new observational paper devoted to *Hα* and Trace UV spicules was recently published by Pasachoff, Jacobson and Sterling, 2009, among many other papers published during the last decade.

The existence of 3-5 min. oscillations in spicules were reported since a long time, for ex. by Kulidzanishvili and Nikolsky (1978). More recently similar periodic oscillations were reported in dark mottles by De Pontieu et al. (2003) and in spicules by Xia et al. (2005), from spectroscopic SUMER data taken with a resolution of 1", and both the De Pontieu et al. and the Xia et al. results seem to agree with each other. Recently Kukhianidze et al. (2006 and 2007) reported periodic spatial distributions of Doppler velocities with height through spectroscopic analysis of $H\alpha$ height series in solar limb spicules (at the heights of 3800-8700 km above the photosphere).

Oscillations in spicules seem to be related to the photospheric $P$-modes oscillations, but it is evident that if spicules oscillations are driven by $P$-modes, crucial details about their formation heights are still missing, because the formation base (feet) of spicules seems to be only 1000 $km$ above the reference photospheric level but they seem not to have any trace at this photospheric level. On the other hand, not all spicules are clearly periodic, whereas most photospheric oscillations are. Spicules have also a rather large aspect ratio, certainly much more than 10, and they occur along a quasi-radial direction, at least in polar regions. These properties suggest **they are confined by the chromospheric magnetic field,** especially where the beta of the plasma becomes small. It makes it difficult to imagine transverse motion of spicules without implying transverse motions of the magnetic field also (Ajabshirizadeh et al. 2009). In addition, the horizontal scale for amplitude coherence of $P$-modes ($\approx$ 8000 km) is well beyond the width of fibrils (De Pontieu et al., 2003 and 2004) and, evidently, off limb spicules.

Spicules are very thin and numerous, so along the line of sight they could be in a large number, especially above the limb where a long integration along the line of sight exists. Then superposition effects (overlapping) should be more important than it was anticipated before, when it was thought that spicules have a 1 Mm or more diameter, because the number of spicules met along the line of sight per resolution element is indeed significant. A kind of collective behaviour of 2 or more components of spicules is not excluded, see Tavabi et al. (2011).

The presence of so-called straws has been described by Rutten (2007) who used observations on the disk with the Dutch Open Telescope in La Palma to describe their morphology and association with the network. Later De Pontieu et al. (2007) carefully describe them using Hinode/SOT data, introducing an additional type of spicules, type II spicules. The ''straws'' like behaviour of spicules (on the limb and especially on the disk) could also be partly due to the superposition effects along the line of sight and this effect should be analyzed.

**2- Observations**

We selected a sequence of solar limb observations made at the North Pole with the broad-band filter instrument (BFI) of the SOT of the Hinode mission (fig. 1, top panel). We use a series of image sequences obtained in 25 October 2008 (near 03h18 U.T.) in the Ca II H emission line; the wavelength pass-band being centered at 398.86 nm with a FWHM of 0.3 nm. A fixed cadence of 8 seconds is used (with an exposure time of 0.5 s) giving a spatial full resolution of the SOT- Hinode (Kosugi et al. 2007 and Tsuneta et al. 2008) limited by the diffraction at $0\overset{''}{.}16$ (120km); a $0\overset{''}{.}0541$ pixel size scale is used.

The size of all images used here is 1024×512 pixels$^2$ (Hinode read out only the central pixels of the larger detector to keep the high cadence within the telemetry restrictions) thus covering an area of (FOV) $111\overset{''}{.}\times56\overset{''}{.}$ at the North Pole as the images are centered at position X=0, Y=948 arcsec. On the polar cap of the Sun spicules are somewhat more numerous than at low latitudes close to the solar equator as they are slightly taller and oriented more radially (Filippov & Koutchmy 2000).

We used the SOT routine program "fg_prep" to reduce the image spikes and jitter effect and to align the time series (Shimizu et al. 2008). The time series show a slow pointing drift, with an average speed less than $0\overset{''}{.}015$ /min toward the north as identified from solar limb motion.

A superior spatial image processing for thread-like features is obtained using the mad-max algorithm (Koutchmy & Koutchmy 1989; Tavabi et al. 2011), see Fig. 1 for ex.

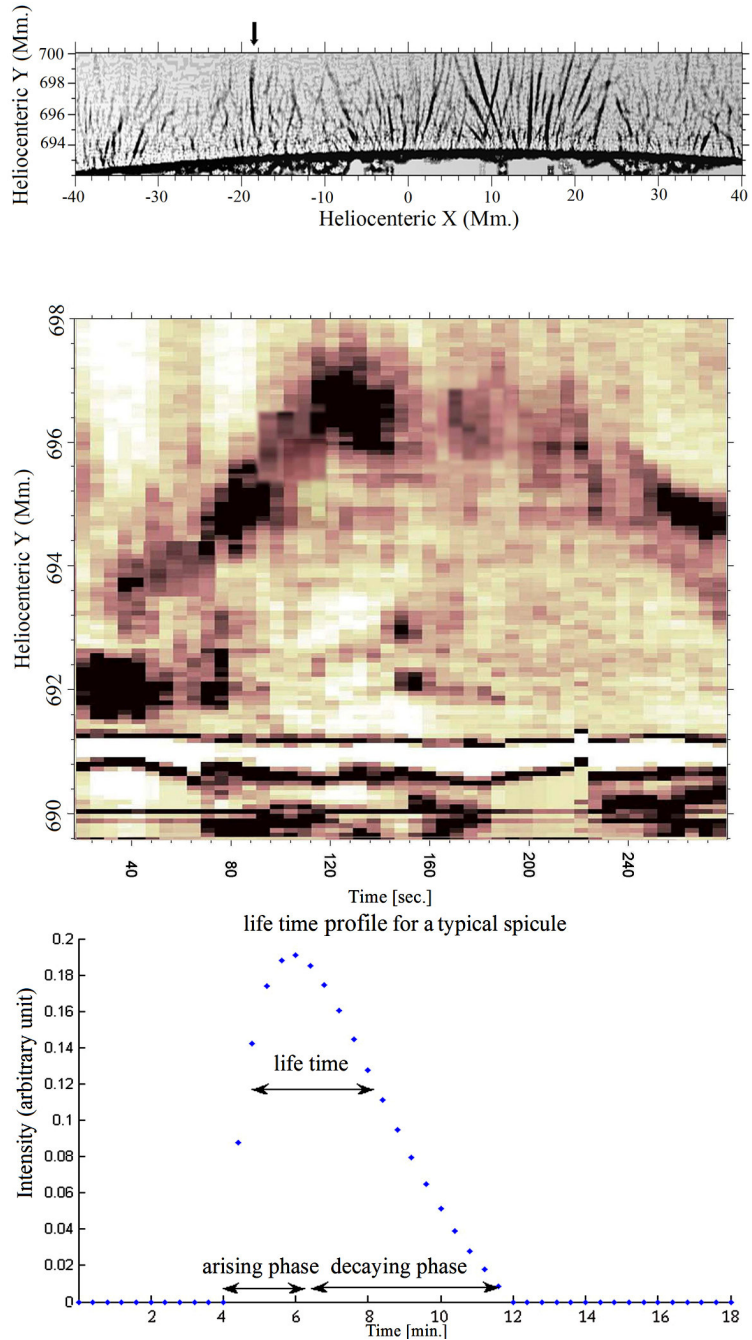

life time profile for a typical spicule

**Fig. 1** Example of a mad-maxed Hinode filtergram in Ca H II, obtained on November 25, 2008, near the North pole (top image), printed in negative. Under is shown a typical Y-t diagram (time-slice diagram) deduced from the analysis of a time sequence of similar images; it gives a ballistic-type motion (see also Suematsu et al. 1995), for the typical spicule shown by a linear mark on the top image, to the left. It is clear

that after reaching a maximum length, the spicules are less rapidly falling back to the solar surface as finally illustrated on the visibility function of the spicule plotted on the bottom panel.

The spatial filtering using the "mad-max" algorithm clearly shows relatively bright radial threads in the chromosphere as fine as the resolution limit of about 120 km, see Fig. 1 and Tavabi et al. 2011, and permits to deduce in 1st approximation, what could be the individual properties of spicules. Note that this rather simple algorithm permits to reduce the complications due to the overlapping effect.

The time-slice diagram in fig. 1 shows a typical upward propagating intensity blob with the apparent propagating speed of about 40 km/s,obtained using high cadence data (8 sec.), this cadence is enough for detecting of very short life-time spicule (type-II), the observed intensity blob may rule out the possibility of fast sausage pulse (Zaqarashvili et al. 2010).

### 3- Simulation

We adopted a preliminary sketch to numerically simulate the limb spicules visibility based on the assumption of a random distribution over the surface.

We know that spicules are located between the chromospheric network of super-granules, the mean distance between being about 20000-30000 $km$, so in this simulation they have been allowed to be in a large number ($n$=100) with a duration about of 60 minutes to approximately match the solar reality. We then analyzed the resulting signal inside an "observational" narrow slit, a slit with a Field-Of-View (FOV) of about 10×30 (pixel)$^2$. The solid line boxes in fig. 2- a, b, c show the slit position for each frame. The time interval between successive frames is 30 $s$. Spicules are assumed to be optically thin such that we add the individual contributions of spicules falling inside the slit.

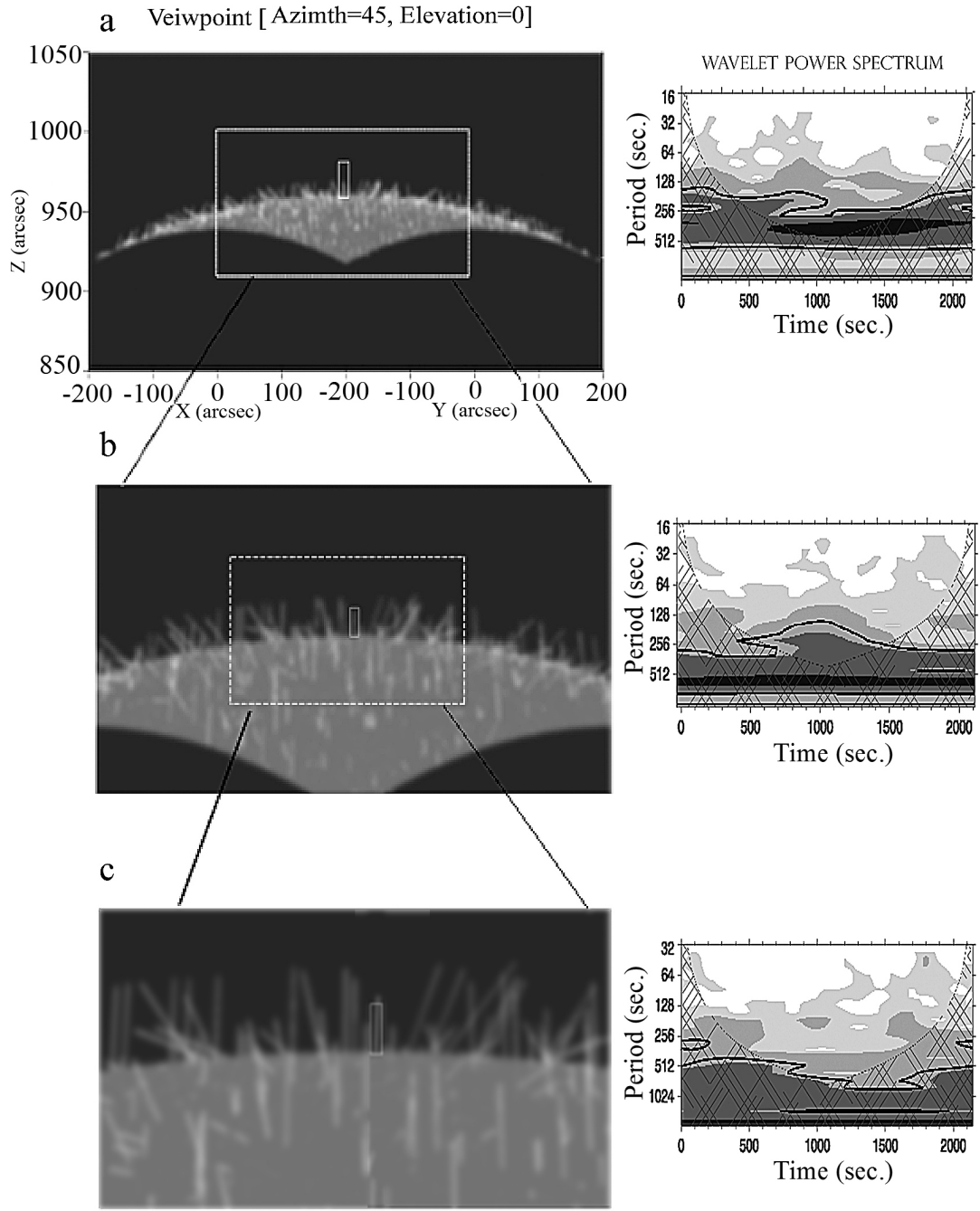

**Fig. 2.** Left frames show the results of simulations for three different FOV (dashed box in upper images is the FOV of the lower images) and small solid boxes are selected area for integration of intensity. The right plots show the corresponding wavelet powers for the integrated intensity fluctuation (period vs. time). Cross-hatched regions indicate the ''cone of influence'' (COI). The darkest regions mean higher power, and the contours correspond to the 95% confidence level (see text and reference for details).

We assume the spicules reach their maximum length after 2-6 minutes and then start to fall back; the arising phase time is shorter than decay phase time and the lifetime for different spicules is varying from 5-15 minutes. In addition, the tilt angle (inclination) of spicules was chosen randomly and increases from pole to the equator, gently (Filippov et al. 2000).

Life-Time: 2-5 min.          *n* = *50*

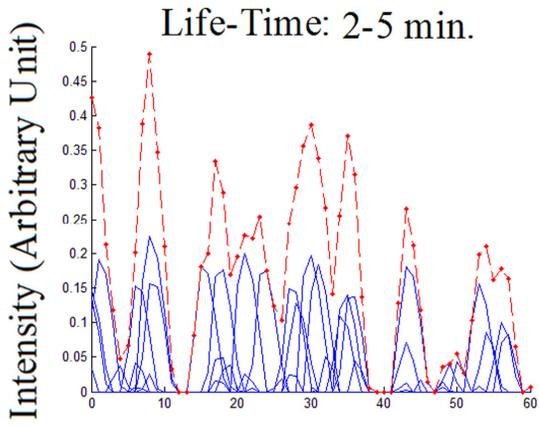

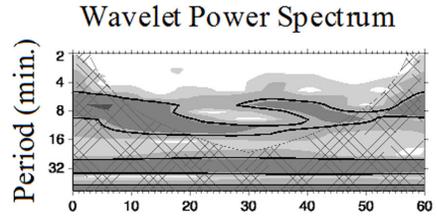

Life-Time:  2-10 min.

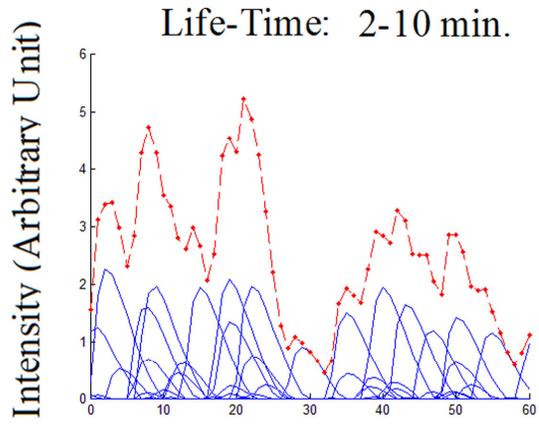

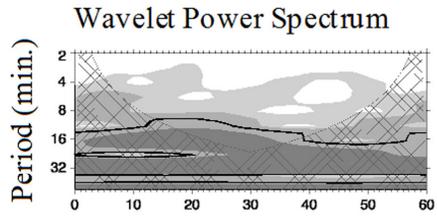

Life-Time:  2-15 min.

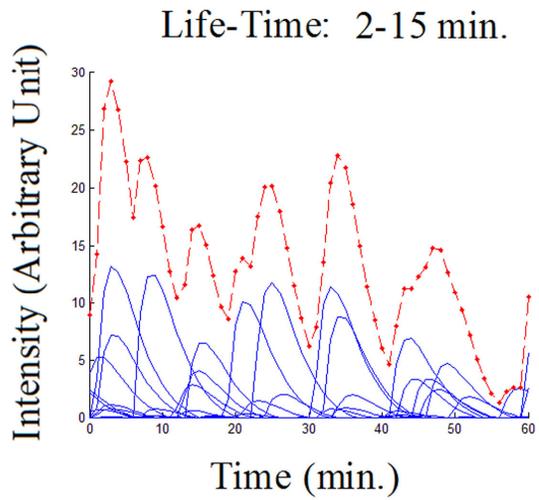

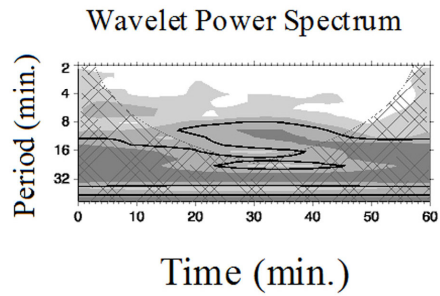

Life-Time: 2-5 min.  *n=100*

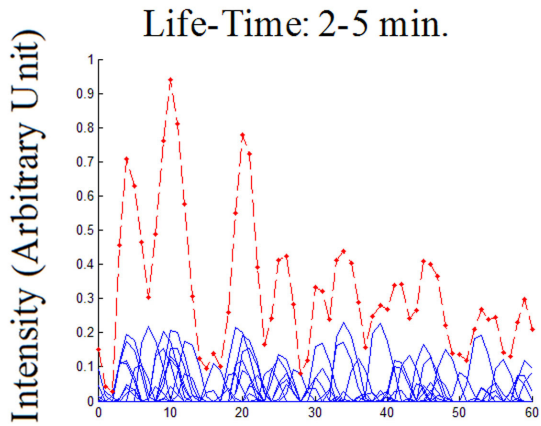

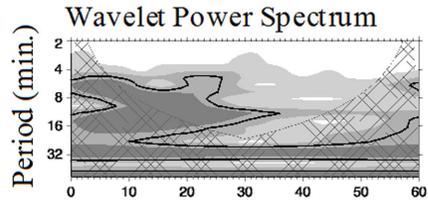

Life-Time: 2-10 min.

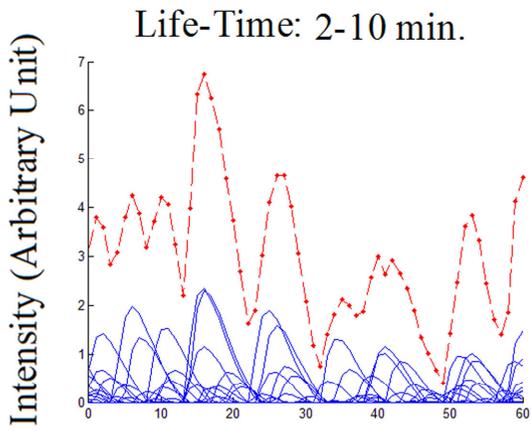

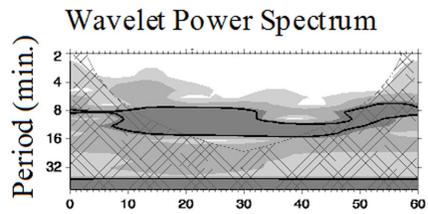

Life-Time: 2-15 min.

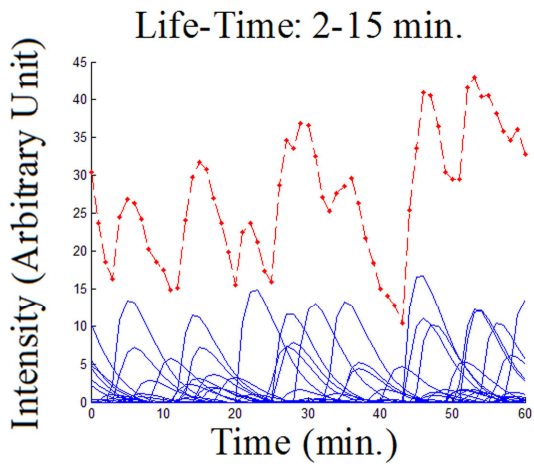

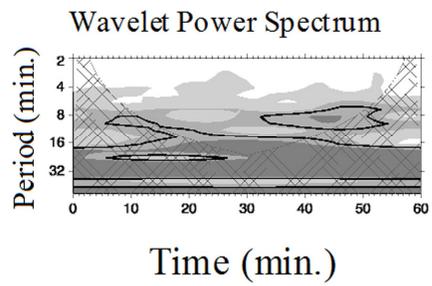

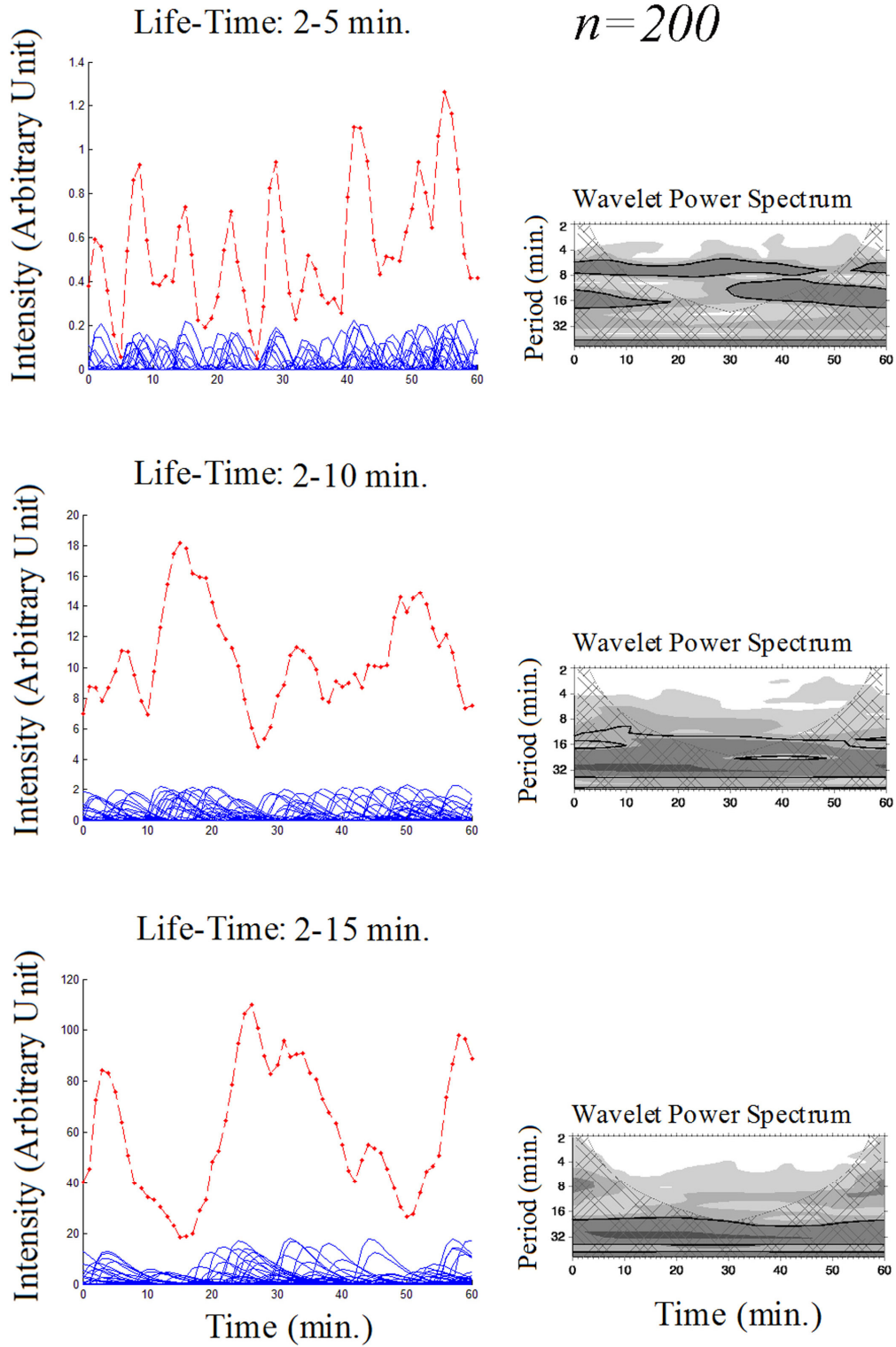

**Fig. 3-a, b & c.** Different light curves resulting from the simulation for different resolutions and their corresponding wavelet spectra. The blue solid curves show the visibility function of spicules for each lifetime interval and the red dotted lines show the envelope of the summed curves. For each envelope, the wavelet power spectra were plotted on the right hand side of the intensity plot. In these figures, we see that the appearance of spicule oscillation is very sensitive to the number of them and to their life-time. With an increased life-time range, the period of the typical oscillation is also increased. The period of oscillations is also sensitive to the number (*n*) of spicules occurring during 60 minutes inside the slit shown in fig. 2, at the middle of the left panels.

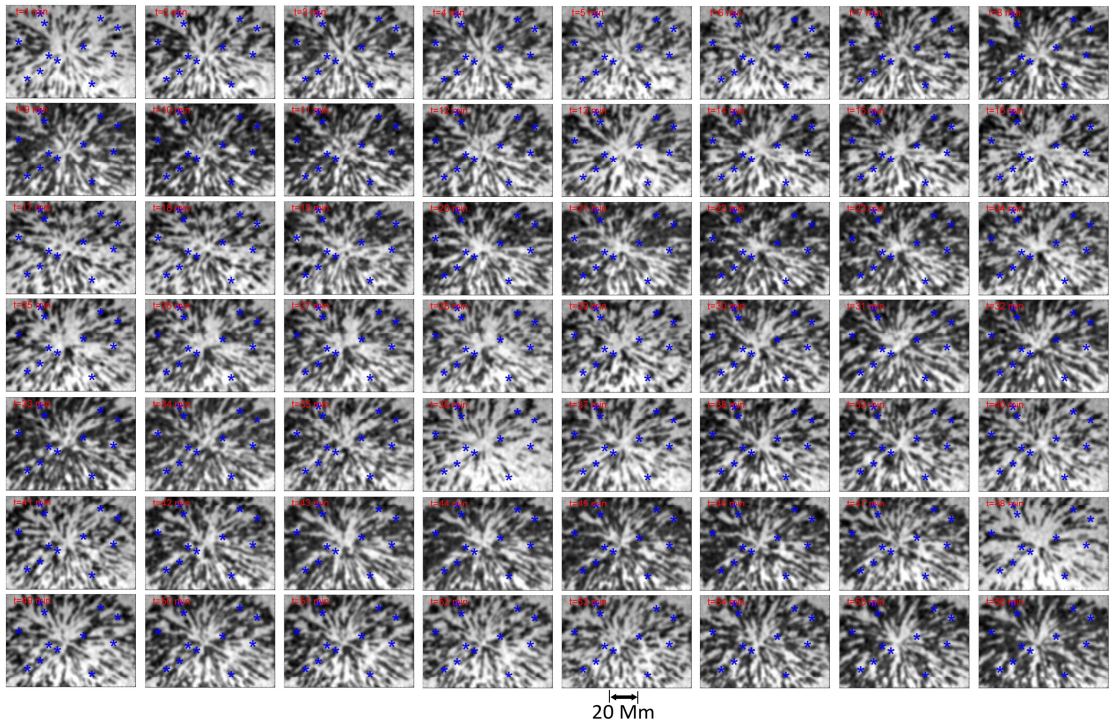

20 Mm

**Fig. 4** Successive frames resulting from a simulation of dark mottles, the lifetime of the mottles being chosen to be about 2~10 minutes. The blue small asterisked position intensity fluctuations have been analyzed by the wavelet time/frequency method, see fig.5. Not all of them, but a large number of the indicated points show a typical period of intensity variation.

Mottles are also jet-like fine quiet Sun structures seen in absorption on the solar disk. They are rather dark features occurring at the boundaries of the chromospheric network. They are classified either into small groups called chains (along the common boundary line of two neighboring granule size cells) or larger groups called rosettes (at the meeting point of at least three network big cells) and they are believed to be the disk counterparts of limb spicules (Tziotziou et al. 2006). Outwardly propagating running waves of 5 min period were reported inside those structures in Doppler velocities, from a rather bright network element observed in 3 different lines, see Baudin et al. (1996). Fast imaging with filtergrams seems to confirm such behavior. It is however well known that solar imaging observations from ground-based telescopes suffer from distortions and smearing; in practice, the seeing in solar observations can be limited by the size of the smallest visible structure in the field of view and it could be comparable with the mottles size.

Like for observation of limb spicules, the tracing of individual mottles seen in projection over the disk is difficult because a large number of them is seen simultaneously. Accordingly, it is generally difficult to track the dark mottles during their lifetime and we met the same identification problem (see fig. 4). In addition, during their lifetime some mottles display a significant fading that could be considered using the classical *cloud* model of chromospheric fine structure (Beckers 1964; Cram 1986). Hence it is very difficult to follow an individual dark mottle. We show the result of our simulation in Fig. 4. Due to the large number of dark mottles with similar life-time and due to the overlapping effect, these frames show an artificial oscillation of dark mottles as illustrated by the wavelet analysis see the wavelet diagrams of Fig. 5. Therefore, it seems that these dynamic structures are also moving up and down with a typical period of recurrence comparable with their lifetime.

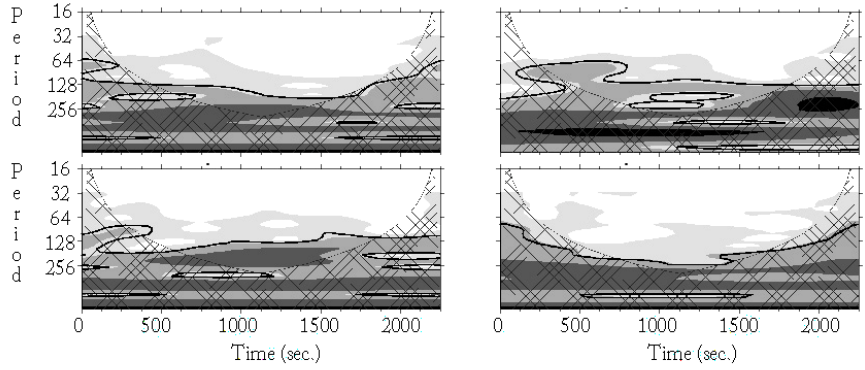

**Fig. 5** Four cases of wavelet power spectrum analysis of intensity fluctuations in dark mottles selected where a blue asterisk is shown in the previous fig. 4. In these plots the 5 minute oscillations clearly appeared for a large part (more than 50 %) of dark mottles.

To fit the observational real situations, we must add seeing and instrumental effects in these simulations; therefore, we used a Gaussian smearing function with a FWHM of about 10 pixels. In fig. 2, the plots show the wavelet power for integrated intensity fluctuation with the duration shown inside the solid line box, as marked in the left frames. In the wavelet power plots: the horizontal axis is the time in seconds, while the vertical axis is the period in seconds. Darker shades correspond to higher power (95% confidence) of the wavelet coefficients. The contours determine the 95% confidence level which was calculated by assuming a white noise background spectrum. Wavelet transform suffers from the ''wraparound'' errors at both edges of a time series (of finite length). The regions in which these effects are important are defined by the cone of influence (COI; see Torrence and Compo, 1998).

In the COI region, the region that suffers from edge effects is chosen as the e-folding time of the Morlet wavelet and it is presented as a hatched area in each plot. From the wavelet time/frequency analysis of this first step of simulation we obtain a typical period of about 250-350s; for the second step of simulation the period is about 800s and for the last step (or very high-resolution frames), we could not find an exact and unique clear value for periods.

**4- Comparison with observations**

Ajabshirizadeh et al. (2008) investigated a series of TRACE images which was taken in the 1600 A$^\circ$ *UV* continuum channel, made on July 6, 2000. As pointed out in their paper, the majority of the chromospheric spicules observed with an unprecedented spatial (0$\overset{..}{}$5) and temporal (15 s) resolution by TRACE shows a periodic behavior with periods about of 3-4 min.

We here present another even more recent result from observations obtained with the Hinode/SOT *Ca II* H-line at 0.16$\overset{..}{}$ spatial resolution (spatial pixel size is about 0.054$\overset{..}{}$), the spatial resolution determined by the diffraction limit of the SOT. The images made on July 12, 2007 in SOT/BFI *Ca II* H 3968 A$^\circ$ with a cadence of 8 *s.* Then we used the same method, the intensity being integrated over an area at the solar limb inside the selected box that covers spicules, and we applied the same wavelet time/frequency analysis. The obtained periods are much larger than what was seen in TRACE results. The periodic behavior of spicules seems irregular and we could not find an accurate value for them (even using a large amount of them, see also De Pontieu et al. 2007 for another discussion). Our result of the wavelet power analysis of spicules is shown in Fig. 6.

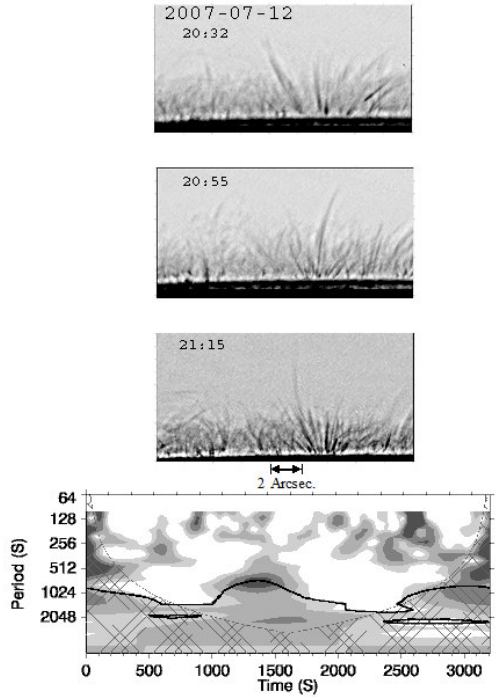

**Fig. 6** Hinode/SOT Ca II-H 3968 A° selected frames of spicules (image scale size is also shown). All images are now real solar images taken from 19:00 to 22:00 UT on 12-July-2007; images are centered at position X=610, Y=710 arcsec in the heliocentric coordinate used to deduce the bottom plot showing the wavelet power and obtain the periods of typical spicule (about of 900s) which was longer than in the case of the TRACE spicules.

The case of non-periodic behavior of spicules has been tested using the very high-resolution images which were taken at Sac-Peak. All images of this set were collected at the prime focus of the Dunn's vacuum Tower Telescope at Sacramento Peak observatory using a digital video CCD camera through a narrow bi-refringent Hα filter and selected afterward. The scale of images is typically 0.048 arcsec/pixel and images were processed using a short burst of 15 images to remove image motion and distortion due to the seeing (see fig. 7). In these images spicules at high latitude but outside the polar region are suddenly ejected from a bottom part very close to the limb and after a few minutes disappear forever, see Lorrain and Koutchmy, 1996. NSO/SPO Observations at the Dunn VTT with high speed Hα line wing imaging reach a typical 200 km resolution,. They show dynamic phenomena like small 1.5 Mm loops at the bottom of the frame and long thin spicules higher up. Recent SOT/Hinode CaII H full line observations show similar phenomena over longer time sequences; more importantly, they are free of seeing effects.

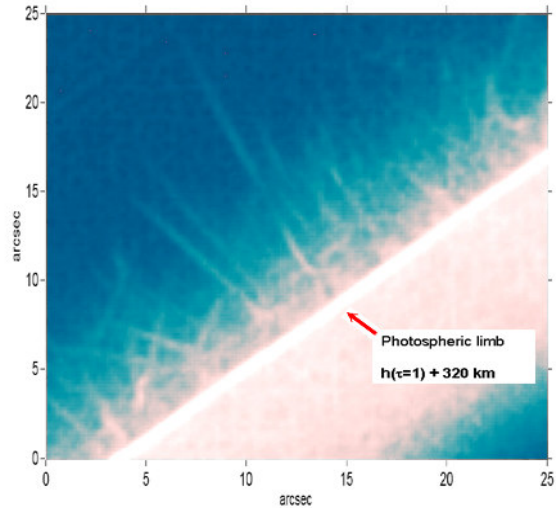

**Fig. 7.** *Very high-resolution image deduced from a burst taken at the Sac-Peak NSO, using the Dunn's vacuum Tower Telescope and the UBF. Images were taken at the video rate using both wings of Hα* **through** *a narrow band Lyot filter and an out-focused mask to avoid over-exposure; the scale of the original images is 0.048 arcsec/pixel, and the field is taken at an intermediate latitude.*

### 5- Summary of results and conclusions.

In fig. 2-a we showed some simulated results over a FOV is $300 \times 300$ (pixel)$^2$ thus covering an area of $200'' \times 200''$; for the following step (fig. 2-b) the FOV of the frames corresponds to a smaller area: $100'' \times 100''$. For the last step (fig. 2-c) the FOV covers only $50'' \times 50''$ on the Sun. Fig. 2 shows that with an increasing resolution of the frames the periods of apparent oscillations increase dramatically. This figure shows that when the resolution is increased (or the pixel size in arcs is decreased), the periods of oscillations increase until the resolution of frames reaches a critical value which was comparable to the spicule true diameter (about 0.2 arcsec) henceforth at this step we could not determine a regular period for spicule recurrence or if a possible period exists its amount is very large. This is a typical result of our simulation.

**Table 1.**

|  | Simulation (first step) | Observation (TRACE 1600A) | Simulation (secont step) | Observation (Hinode CaII H) | Simulation (third step) | Observation (Sac-peak H$\alpha$) |
|---|---|---|---|---|---|---|
| Resolution (arcsec) | 0.44 | 0.5 | 0.11 | 0.16 | 0.027 | 0.048 |
| Period (S) | 300 | 280 | 850 | 900 | None | None |

All evidence presented so far is outlined in table 1 where the simulation results are compared to the observations for similar resolution; the periods for each corresponding resolution are showed to be similar (see also fig. 3).

We further considered some space and ground-based spicules observations in three different optical lines which were approximately comparable to the resolution of three steps simulations respectively. The results of the simulation seem to fit these observations. It is then possible to conclude that the superposition (overlapping) of many finely-structured features undergoing vigorous longitudinal spicular motions naturally leads to the observed periodic features of spicules. Spicules are very thin so the line of sight could be occupied by a myriad of them. In addition many spicules have a typical lifetime near 5 minutes and they occurred repeatedly. So, we "see" them as an oscillatory structure. It is quite possible that observations with a resolution of order of two to three times coarser than the individual spicule diameter missed most of the dynamical behavior of the spicules.

Additionally, some small loops were identified in fig. 2-a; but when we look at the higher resolution frames we seemingly recognize that these loops could be just due to the overlapping of several neighbored spicules with different tilt angle with respect to each other. This point will have to be considered in a forthcoming paper.

**Acknowledgements.** The authors thank B. Filippov for a critical reading of the manuscript. We are grateful to the Hinode team for their wonderful observations. Hinode is a Japanese mission developed and launched by ISAS/JAXA, with NAOJ as domestic partner and NASA, ESA and STFC (UK) as international partners. TRACE is a mission supported by NASA and Lockheed. Image processing wavelet software was provided by O. Koutchmy http://www.ann.jussieu.fr/~koutchmy/index_newE.html, and wavelet analysis software was provided by Torrence and Compo (http://atoc.colorado.edu/research/wavelets). This work was partly financially supported by RIAAM (Iran) and the French CNRS.